\documentclass{iopart}
\usepackage[english]{babel}
\usepackage[utf8x]{inputenc}
\usepackage[T1]{fontenc}
\usepackage{amssymb}
\usepackage[dvipdfm]{graphicx}

\newcommand{\ft}{f_\mathrm{T}}

\newcommand{\TT}{\left[\mathrm{TT}\right]}
\newcommand{\TL}{\left[\mathrm{TL}\right]}
\newcommand{\LL}{\left[\mathrm{LL}\right]}
\newcommand{\TTT}{\left[\mathrm{TTT}\right]}
\newcommand{\TTTt}{{\left[\mathrm{TTT}\right]}_\vartriangle}
\newcommand{\TLTt}{{\left[\mathrm{TLT}\right]}_\vartriangle}
\newcommand{\TLL}{\left[\mathrm{TLL}\right]}
\newcommand{\TTL}{\left[\mathrm{TTL}\right]}
\newcommand{\TLLt}{{\left[\mathrm{TLL}\right]}_\vartriangle}
\newcommand{\TTLt}{{\left[\mathrm{TTL}\right]}_\vartriangle}
\newcommand{\LLL}{\left[\mathrm{LLL}\right]}
\newcommand{\LTT}{\left[\mathrm{LTT}\right]}
\newcommand{\LTLt}{{\left[\mathrm{LTL}\right]}_\vartriangle}

\newcommand{\KL}{{\left\langle k\right\rangle}_\mathrm{L}}

\newcommand{\pA}{p_\mathrm{A}}
\newcommand{\pT}{p_\mathrm{T}}
\newcommand{\pL}{p_\mathrm{L}}
\newcommand{\binom}[2]{\left(\begin{array}{c}{#1}\\{#2}\end{array}\right)}

\begin{document}
\title[Emergent bipartiteness in a society of knights and knaves]
{Emergent bipartiteness in a society of knights and knaves}

\author{C I Del Genio$^1$ and T Gross$^{1,2}$}
\address{$1$ Max-Planck-Institut für Physik komplexer Systeme, Nöthnitzer Straße 38, 01187 Dresden, Germany}
\address{$2$ Department of Engineering Mathematics, University of Bristol, Merchant Venturers Building, Bristol BS8 1UB, United Kingdom}
\ead{paraw@pks.mpg.de}
\begin{abstract}
We propose a simple model of a social network based on so-called knights-and-knaves puzzles.
The model describes the formation of networks between two classes of agents
where links are formed by agents introducing their neighbors to others of their own class. 
We show that if the proportion of knights and knaves is within a certain range, 
the network self-organizes to a perfectly bipartite state. 
However, if the excess of one of the two classes is greater than a threshold value, bipartiteness is not observed. 
We offer a detailed theoretical analysis for the behaviour of the model, investigate
its behavior in the thermodynamic limit, and argue that it provides a simple example of a topology-driven
model whose behaviour is strongly reminiscent of a first-order phase transitions far from equilibrium.
\end{abstract}
\submitto{\NJP}
\maketitle

\section{Introduction}
\emph{``The only way out of here is to try one of these doors. One of them leads to the castle
at the centre of the Labyrinth, and the other one leads to certain death! You can only ask one
of us, and I should warn you that one of us always tells the truth, and the other always lies.''}~\cite{LabMOV}
The statement above poses a Knights-and-Knaves puzzle -- a class of logic puzzles
made popular by Raymond Smullyan~\cite{Smu09}. As their defining feature, these puzzles contain
two types of characters: the knights, who always tell the truth, and the knaves, who always lie.

In physics and mathematics, the investigation of simple puzzles and toy models has often led to
deep insights. For instance, puzzles of the Knights-and-Knaves type are quoted as an inspiration
for Gödel's incompleteness theorem~\cite{Goe31}. Examples of influential simple models from physics
include the Ising model~\cite{Isi25} and the Bak-Tang-Wiesenfeld model of self-organized criticality~\cite{Bak87}.
In the physics of complex networks, simple models have significantly advanced our understanding
of both the topological evolution of networks~\cite{Pri65,Bar99} and the dynamical processes taking place
on them~\cite{Kau69,Hop82}. More recently, studies of the adaptive voter model~\cite{Nar08,Ser09},
a highly simplified model of opinion dynamics, have resulted in a better understanding of adaptive
networks, which are networks in which the topology coevolves with the state of the nodes~\cite{Gro07}.

Here, we propose a very simple network formation game~\cite{Bal00,Do10}, inspired by Knights and Knaves puzzles.
In contrast to traditional puzzles, the model (described in \sref{secModel})
considers the dynamics of a social network of knights and knaves. We assume that
every agent, regardless of his own character, tries to connect to knights while
avoiding knaves. However, by nature of this game, every agent will claim to be a knight
if asked directly. Therefore, the agents have to rely on social information,
asking their neighbours with whom to link and whom to avoid.

One of the solutions to the puzzle posed above is to ask one of the agents
which door the other agent would recommend. The agent will then invariably
name the door that leads to death, thus implicitly revealing the door that
leads to the castle. This solution exploits a symmetry of the puzzle: a knight 
relating the answer of a knave will result in the same information as a knave relating 
the answer of a knight. Alternatively, one can ask one of agents what he 
would recommend if asked directly. A knight will truthfully relate his true answer, 
while a knave will lie about his lie; in both cases the right door is named.

The symmetry of the knights-and-knaves puzzles carries over to the proposed network 
model. A knave will always recommend linking to knaves, pretending them to be knights. By contrast, a
knight will always recommend linking to knights, truthfully revealing their
knightly character. Thus, a symmetric situation arises in which every agent
recommends those of his own type.

Based on the above, one might argue that the model has some significance for opinion formation processes,
with, e.g., Republicans referring their discussion partners to other Republicans and Democrats referring
to other Democrats. However, our main motivation for studying the network of knights and knaves stems from a
different source: The model proposed here is one of the simplest nonlocal systems exhibiting nontrivial topological
dynamics. Thus, it constitutes a step toward the exploration of mesoscale dynamics in networks.

In this paper we study the dynamics of the Knights-and-Knaves network numerically and analytically.
One question that immediately comes to mind is whether the knights manage to separate themselves from the knaves.
In a wide range of parameters, the opposite turns out to be true: The network approaches a completely
bipartite state in which every knight is connected only to knaves and every knave is connected only
to knights. Bipartiteness is still achieved if there is a significant difference in the numbers
of knights and knaves, but disappears when the difference exceeds a certain threshold. Our analysis
reveals a strong analogy between the behaviour of the system and thermodynamic properties close to
first order phase transitions. The proposed model may thus offer an analytically tractable example
of such a transition in a topology-driven finite-temperature system far from equilibrium.

\section{The model}\label{secModel}
We consider a network of $T$ knights (T for truthful)
and $L$ knaves (L for liar), such that the total number of nodes
is $N=T+L$ and the proportion of knights in the population is $\ft=T/N$. 

The network starts from some random initial configuration and then evolves 
according to the following rules: In every time
step we randomly chose a node, $i$, one of its neighbors, $j$, and
one of $j$'s neighbors, $k\neq i$. If node $j$ and $k$ are of identical
type (both T or both L) then $i$ connects to $k$ or maintains the
connection to $k$ if one exists already. If node $j$ and $k$ are
of different type (one T, one L) then $i$ does not connect to $k$
and cuts the connection to $k$ if one exists already. This procedure
is iterated until the system reaches either an absorbing state,
where no further change of the topology is possible, or a thermodynamic
steady state, in which the microscopic dynamics continues.

Similar simple models have also been discussed in the context of balance theory~\cite{Ant05,Ant06}.
However, where balance models focus on links of two different kinds, with no difference amongst nodes,
the model proposed here considers nodes belonging to two different classes, with no distinction amongst
links. Also, the dynamics we defined continuously changes the topology of the network, which is instead
mantained unchanged in balance theory.

For the analysis below it is useful to define $\TT$ as the total number of links between
knights, $\LL$ as the total number of links between knaves, and $\TL$ as the
total number of links between a knight and a knave.

\section{Numerical results}

\begin{figure}
 \centering
\includegraphics[width=0.6\textwidth]{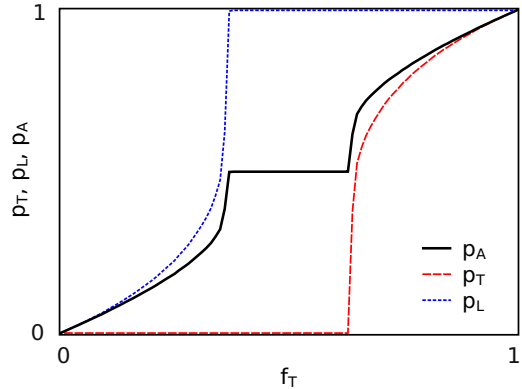}
\caption{\label{Fig1}Emergent bipartiteness. Ensemble averaged probabilities for a neighbour of a knight to be a knight
($\pT$, dashed red line), for a neighbour of a knave to be a knight ($\pL$, dotted blue line), and for
a neighbour of any node to be a knight ($\pA$, solid black line), as a function of the fraction $\ft$
of knights in the network. The lines are averages over an ensemble of $10^4$ networks with $N=10^4$ nodes.
The networks exhibit perfect bipartiteness in the region roughly between $\ft=0.37$ and $\ft=0.63$.}
\end{figure}

For investigating the phenomenology of the model we start by defining thermodynamic observables.
Because we motivated the model by assuming that the agents aim to connect to knights, it is reasonable to introduce
observables that measure how well this goal is achieved. We measure the success of knights by a
parameter $\pT=2\TT/\left(2\TT+\TL\right)$, denoting the proportion of neighbors of knights that
are knights. Analogously, we define $\pL=\TL/\left(\TL+2\LL\right)$ as the proportion of neighbors
of knaves that are knights. Finally, we denote the probability that a knight is reached by following
a random link as $\pA=\left(2\TT+\TL\right)/\left[2\left(\TT+\TL+\LL\right)\right]$.

In simulations, we observe that if the number of knights equals the number of knaves
then the system always evolves to a state where all neighbors of knights are knaves
and all neighbors of knaves are knights, i.e., $\pT=0$, $\pL=1$ and $\pA=0.5$. In the
following we denote this state as the bipartite state of the network.

Once in the bipartite state the dynamics freezes: Given two nodes $i$ and $k$,
with common neighbour $j$, either $i$ and $k$ are knights while $j$ is a knave,
or $i$ and $k$ are knaves while $j$ is a knight. Either way, $j$ is of a type
different from both $i$ and $k$. Hence, no link between $i$ and $k$ can be placed and 
no link from $i$ to $k$ can exists, which could be removed.

We now ask whether the bipartite state can still be reached if the proportion
of knights and knaves in the population is different. Simulations of the network
dynamics for different values of $N$ and $\ft$ show that the equilibrium network
is bipartite throughout a range of values of $\ft$ centered around $0.5$ (\fref{Fig1}),
whereas bipartiteness is lost if the proportion of knights or knaves exceeds
some threshold.  Thus, depending on the value of $\ft$ three different regimes
are observed: First, at low $\ft$ the knights are exclusively connected to knaves,
whereas knaves have additional connections among themselves. Second, at $\ft$
around 0.5 both knights and knaves are exclusively connected to agents of the other
type (bipartite state). Third, at high $\ft$ the knaves are exlusively connected
to knights, whereas the knights also have connections among themselves.

\begin{figure}
 \centering
\includegraphics[width=0.6\textwidth]{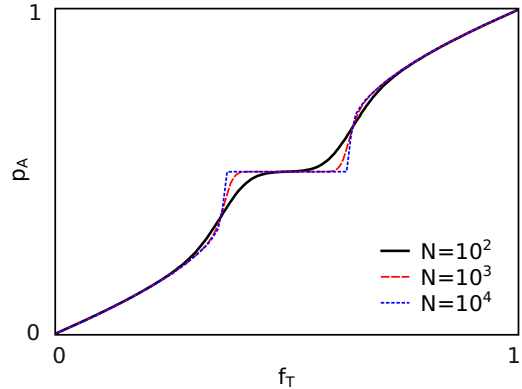}
\caption{\label{Fig2}Effect of system size. Shown is the ensemble averaged probability for the neighbour
of a node to be a knight ($\pA$), as a function of the fraction $\ft$ of knights in the network, for different
system sizes. The solid black line corresponds to $N=10^2$, the dashed red line to $N=10^3$, and the dotted
blue line to $N=10^4$. The bipartiteness range decreases with system size, and vanishes for $N=10^2$.
The lines for $N=10^3$ and $N=10^4$ are averages over ensembles of $10^4$
networks; the line for $N=10^2$ is an average over an ensemble of $10^6$ networks.}
\end{figure}

Let us emphasize that the behaviour of $\pA$ is strongly reminiscent of the Maxwell construction for the isotherms
of the van~der~Waals equation, or the M-H isotherms of magnetic systems undergoing first-order phase transitions~\cite{Sta71}.
In this context of phase transitions, the model should be considered as a systems at non-zero temperature. While we
prescribed the result of the update of given triplet deterministically, the triplets to update are chosen randomly.
This stochasticity constitutes a finite temperature, which is higher in networks of smaller size. We note that phase
transitions can only be strictly defined for systems of infinite size. However, the space of possible topologies of
networks scales with $2^{N^2}$, where $N$ is the number of network nodes. Therefore, even relatively small networks
constitute a large configurational space, meriting the application of statistical concepts.

In \fref{Fig2}, the value of $\pA$ in the final state of the network is shown for different network sizes.
We observe that the range of bipartiteness decreases with decreasing system size. If the network size is shrunk
down to $N=100$ nodes then the range of bipartiteness becomes a single point reminiscent of the critical point
of the van~der~Waals equation. While this analogy should certainly explored in more detail by artificially introducing
noise in larger networks, this investigation is beyond the scope of the current paper. Instead, we focus on the dynamical
origin of the bipartite regime.

\section{Thermodynamic theory}
\begin{figure}
 \centering
\includegraphics[width=0.5\textwidth]{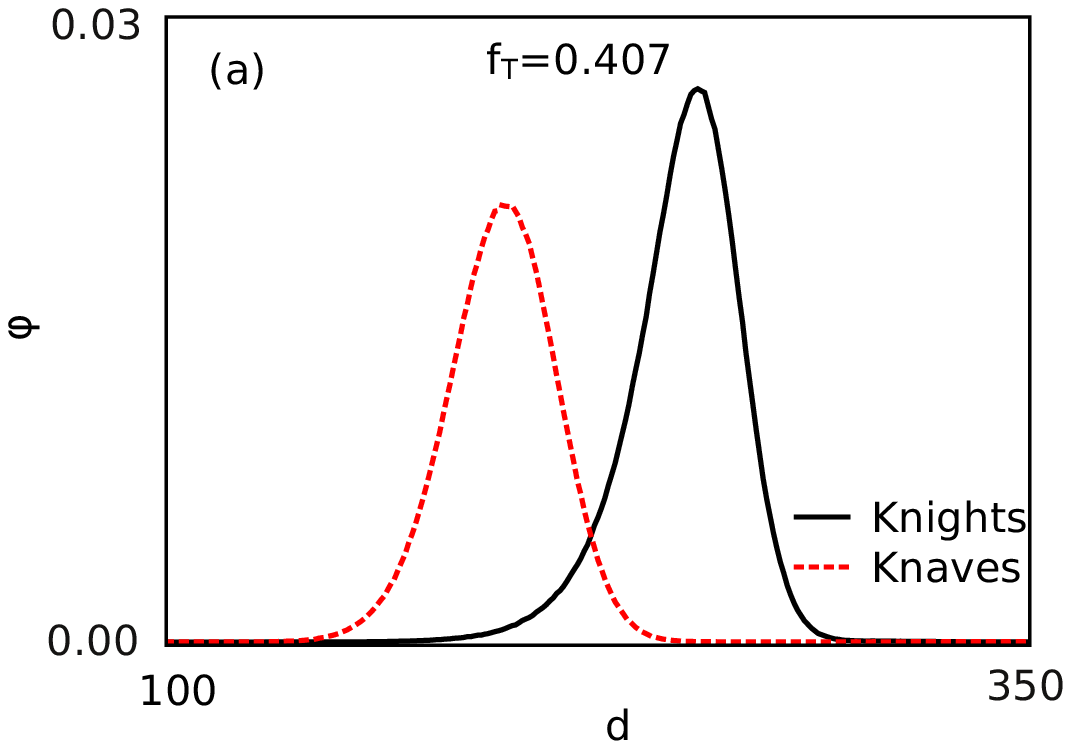}
\includegraphics[width=0.5\textwidth]{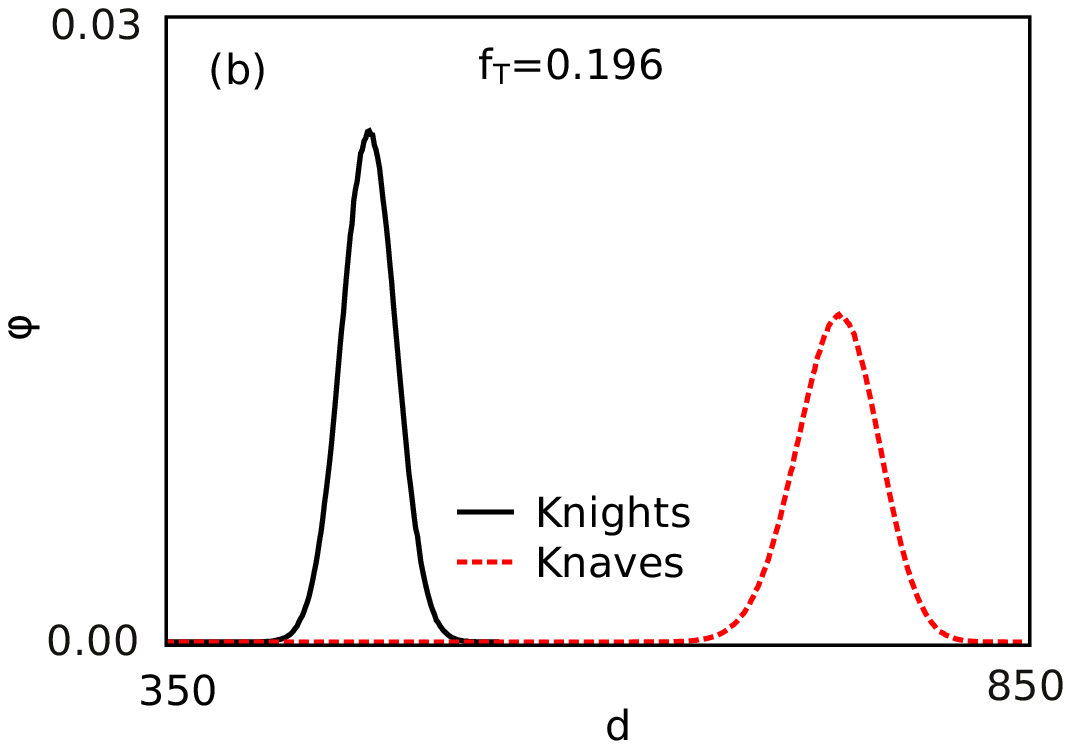}
\caption{\label{Fig3}Ensemble averaged degree distributions $\varphi$
for knights (solid black line) and knaves (dotted red line), with $N=10^3$.
The lines are averages over an ensemble of $10^4$ networks. Panel (a)
shows data for $\ft=0.407$: the distribution of the majority nodes (knaves)
``shifts'' towards lower degrees, while that of the minority nodes (knights)
shifts towards higher degrees. Panel (b) shows data for $\ft=0.196$,
outside the bipartiteness range: the minority nodes (knights) have consistenly
lower degrees than the majority ones (knaves).}
\end{figure}
The regimes described in the previous section are ``thermodynamic'', in the sense that they present
very probable, but not strictly certain, outcomes of the network evolution. To see this consider
for instance a network containing only one knight among a large number of knaves, corresponding to a
value of $\ft$ well below the observed bipartite range. Even in this case it is still possible
to construct a bipartite network configuration in which the knight is connected to every other agent
and no further links exist in the system, such that the network is in an absorbing bipartite state.
However, the probability that this configuration arises in the evolution of the network (in finite time)
is so low that it is never observed in any network with more than a few nodes.

While in very small systems any of the three macro-states (bipartite network -- connections between knights but
not knaves -- connections between knaves but not knights) could be observed with some probability,
we observe that larger networks reliably select one of the three behaviors depending the control parameter
$\ft$.

For understanding the mechanism behind this selection it is instructive to consider the distribution
of \emph{degrees} (i.e., the number of connections) for the agents of the two types.
The results presented above show that, within a certain range, the probability
that the neighbor of a random agent is a knight is $0.5$ regardless of the precise
density of knights in the network. This is possible because the agents of the type
numerically in excess have a proportionally lower number of network connections
per agent (see \fref{Fig3}). However, if the proportion of knights or knaves becomes
too small, the agents in the majority start forming connections among each other.
The transitions bordering the bipartite regime can thus be understood as the nucleation
of ``droplets'' of connected majority nodes from the bipartite mixture.

In the remainder of this paper we investigate the formation and breakdown of the bipartite state.
Let us first motivate the existence of this state by a thermodynamic argument, in which, for the
moment, we forget our knowledge of the microscopic dynamics. It is clear that the mean degree of
agents is strongly controlled by the microscopic rules and thus cannot be inferred from a purely
thermodynamic perspective. This situation is similar to a thermodynamic system whose energy is controlled
by coupling to an external heat bath. Therefore, we consider an ensemble of systems with a given
mean degree, which is somewhat analogous to the micro-canonical ensemble of equilibrium statistical
physics.

In the present non-equilibrium system there is no reason to believe that the microstates
in the ensemble should be equiprobable. We nevertheless draw on the micro-canoncial picture
and adopt equiprobability as an admittedly naive working hypothesis. Under this hypothesis
one can then argue that one should observe the macro-state corresponding to the largest
number of micro-states. In other words, we expect to observe the bipartite state when the
number of micro-states that are bipartite is greater than the number of micro-states in
which connections between nodes of the same type exist.

In a network model a microstate corresponds to a distinct realization of the network topology,
e.g.~the precise pattern of neighborhood relationships. We could now proceed by computing closed
expressions for the respective numbers of micro-states and asking at which value of $\ft$ the
micro-states corresponding to bipartiteness are in the majority. While we will indeed derive similar
expressions below, let us first follow a simpler approach leading to the same result. We estimate
the relative number of configurations by considering the bipartite state and comparing the number
of bipartite and non-bipartite configurations that are reached by rewiring one link.

Without loss of generality we assume that the knaves are in the majority. In this case placing a link
between two knaves clearly leads to a larger number configurations than placing a link between two knights.
Given that $\TL$ bipartite links already exist in the network, the number of possibilities for placing
the rewired link between a knight and a knave is
\begin{equation*}
Q_{\mathrm{TL}} = TL-\TL\:,
\end{equation*}
whereas the number of possibilities for placing a link between two knaves is
\begin{equation*}
Q_{\mathrm{LL}} = \frac{L\left(L-1\right)}{2}-\LL\:.
\end{equation*}
The number of admissible bipartite configurations exceeds the number of admissible
non-bipartite configurations when
\begin{equation}\label{eqpz}
Q_{\mathrm{TL}} > Q_{\mathrm{LL}}\:,
\end{equation}
or, equivalently,
\begin{equation*}
 TL-\TL > \frac{L\left(L-1\right)}{2}-\LL\:.
\end{equation*}
If no links between two knaves are ever placed, $\LL=0$. Then, for large $L$,
dividing the above inequality by $L$, we get
\begin{equation*}
 T-\KL>\frac{L}{2}\:,
\end{equation*}
where we used $\frac{\TL}{L}=\KL$. Replacing $T$ with $N-L$ and solving for $L$, yields
\begin{equation}\label{eqCond}
 L<\frac{2}{3}\left(N-\KL\right)\:.
\end{equation}

Following the reasoning above, we would expect to observe the bipartite state
whenever the condition in Eq.~\eref{eqCond} is met. The simple thermodynamic
reasoning therefore predicts the observation of the bipartite state if the
difference in the proportion of minority and majority nodes is sufficiently small.
For gaining a quantitative estimate of the transition point we assume
$\KL\approx N/7$, which we observed in network simulations, independently of $N$.
This yields a bipartite range of $0.43\lessapprox\ft\lessapprox 0.57$.

The results in Fig.~\fref{Fig4} show that the thermodynamic estimate of the
transition point is of the right order of magnitude but differs by some percent
from the value observed in network simulations. The discrepancy can be attributed to
the simplicity of the thermodynamic estimation, which used the unwarranted assumption
of equiprobability of states and neglected the microscopic dynamics. For obtaining
a more precise estimate we formulate a microscopic kinetic theory in the following section.

\section{Kinetic theory}
In this section we use a a network moment expansion~\cite{Kee97,Gro08}
to formulate a set of coarse-gained equations that capture the emergent-level
dynamics of the system. Considering the effect of link creation and destruction
processes on the abundance of links of a given type leads to the equations
\begin{eqnarray}
 \frac{\rmd}{\rmd \mathrm t}\TT &= 2\TTT-2\TLTt\label{MCA1}\\
 \frac{\rmd}{\rmd \mathrm t}\TL &= \TLL+\TTL-2\TLLt-2\TTLt\label{MCA2}\\
 \frac{\rmd}{\rmd \mathrm t}\LL &= 2\LLL-2\LTLt\label{MCA3}\:.
\end{eqnarray}
where we used three-letter symbols indicate open triplets of nodes and triangles.
For example, $\TTT$ indicates the number of open-chain triplets made of three knights,
while $\TLTt$ refers to the number of triangles composed of two knights and one knave.
The symbols $\TLL$, $\TTL$, $\TLLt$, $\TTLt$, $\LLL$, $\LTT$, and $\LTLt$ are defined
analogously.

Because of the appearance of three-node motifs, the equations above do not constitute a closed system.
We close the system by the so-called moment closure approximation~\cite{Kee97,Gro06}, which
replaces the abundances of three node motifs by a statistical estimate based on the abundances of smaller
motifs. Thus, we express the number of triplets as given by the possibilities one has of picking
its two constituent couples with the constraint that they share a given node. So, for instance,
to form a $\TLL$ triplet we start by taking a TL-couple. Each of the further links the knave
of the couple forms is an LL-couple with probability proportional to $2\LL/L$. Since we are
interested in \emph{open} triplets, we have to subtract the number of TLL-triangles, which
we obtain with a similar argument. This leads to the set of moment-closure equations
\begin{equation}\label{eqTTT}
 \TTT=\frac{2\TT^2}{T} - \frac{4\TT^3}{T^3}\:,
\end{equation}
\begin{equation}\label{eqTTTt}
 \TTTt=\frac{4\TT^3}{T^3}
\end{equation}
\begin{equation}\label{eqTLL}
 \TLL=\frac{2\TL\LL}{L} - \frac{\TL^2\LL}{TL^2}\:,
\end{equation}
\begin{equation}\label{eqTLLt}
 \TLLt = \frac{\TL^2\LL}{TL^2}\:,
\end{equation}
where we used the observation that the degree distribution is sufficiently narrow to be treated as possonian
and assumed the absence of correlations beyond the next neighbor. The remaining expressions are easily obtained
from~\eref{eqTTT}, \eref{eqTTTt}, \eref{eqTLL} and~\eref{eqTLLt} exploiting the symmetry of the system, the
invariance of triangles under permutation of their nodes, and the invariance of open triplets under exchange
of their end nodes.

Applying the moment closure approximation to \eref{MCA1}, \eref{MCA2} and \eref{MCA3}, we obtain
\begin{equation}\label{MCAsystem}
 \eqalign{\frac{\rmd}{\rmd \mathrm t}\TT &= \frac{2\TT}{T}\left(2\TT - \frac{4\TT^2}{T^2}-\frac{\TL^2}{TL}\right)\\
 \frac{\rmd}{\rmd \mathrm t}\TL &= \TL\left(\frac{\TT}{T}+\frac{\LL}{L}\right)\left(2-\frac{3\TL}{TL}\right)\\
 \frac{\rmd}{\rmd \mathrm t}\LL &= \frac{2\LL}{L}\left(2\LL-\frac{4\LL^2}{L^2}-\frac{\TL^2}{TL}\right)\:.}
\end{equation}

\begin{figure}
 \centering
\includegraphics[width=0.6\textwidth]{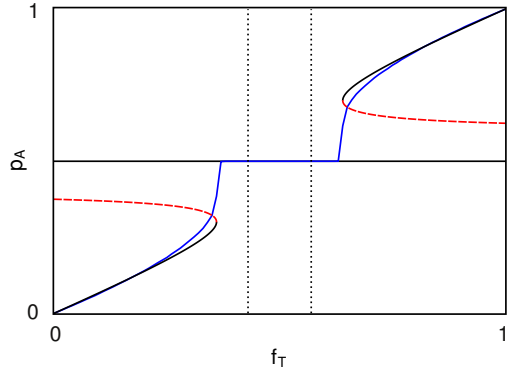}
\caption{\label{Fig4} Comparison of the transitions points.
The figure shows the stationary probabilities for neighbours
of any node to be knights ($\pA$), estimated via moment closure
approximation (black and red lines) and by direct network simulation
with $N=10^4$ (blue line). Furthermore, transition points obtained
from a simple ``thermodynamic'' estimate are shown (dashed
grey lines). The analytical moment-closure approximation offers
a very precise estimate of the transition points. In the approximation
these points correspond to saddle-node bifurcations, where
a stable (black) and an unstable (red) steady state collide
and annihilate.}
\end{figure}

The equation system, Eqs.~\eref{MCAsystem}, is analytically tractable.
We analyze the system by computation of stationary states and a subsequent
linear stability and bifurcation analysis. The bipartite state $\TT=\LL=0$
is trivially stationary. In the range where bipartiteness is actually
observed, this state is the only attractor of the system. If $\ft$ is
increased or decreased beyond the bipartite regime, a qualitative transition
of the dynamics is encountered where two additional stationary states
are formed (\fref{Fig4}), one of which is dynamically stable. We can identify
these transitions as saddle-node bifurcations (also called fold bifurcations
in the mathematical literature).

The results in \fref{Fig4} show that the saddle-node bifurcations
coincide very well with the transition points observed in the network
simulations. We observe a small discrepancy between the estimated
and observed quantities close to the border of the bipartite regime.
Notably, the transitions bordering the bipartite regime look continuous
in the network simulations but are discontinuous in the analytical
approximation. These differences arise most probably because of finite
size effects in the network simulation or because of the presence
of long-ranged correlations which are neglected in the moment-closure
approximation.

Our main conclusion from the thermodynamic plausibility argument
and the analytical model is that the qualitatively different regimes
persist in the thermodynamic limit of infinite network size. We
perceive this observation as a strong encouragement for considering
the observed phenomenon as a phase transition.

\section{State selection}
The analytical approximation showed that within the bipartite regime,
the bipartite state is the only attractor of the system. However, outside
the bipartite regime a stable steady state coexists with the bipartite
absorbing state. While network simulation showed that the system always
approaches the non-bipartite state in this case, the same information
cannot be obtained analytically from the differential equations alone.
We explore this point further by combining results from the analytical
approximation with the thermodynamic reasoning used above.

A drawback of our thermodynamic arguments was that we had to use heuristic
values for the density of links between given types of agents. This drawback
can now be mitigated by using results from the moment-closure approximation.
We start by writing the number of possible configurations $S$ for a
network with a given proportion of knights in each of the states, which yields
\begin{equation*}
 S = \binom{TL}{\TL}\binom{\frac{T\left(T-1\right)}{2}}{\TT}\binom{\frac{L\left(L-1\right)}{2}}{\LL}\:,
\end{equation*}
where $\binom{a}{b}=a!/\left[b!\left(a-b\right)!\right]$ is the binomial coefficient. In this equation the three factors
arise form the number of possibilities for placing the TT, TL, and LL links, respectively.

\begin{figure}
 \centering
\includegraphics[width=0.6\textwidth]{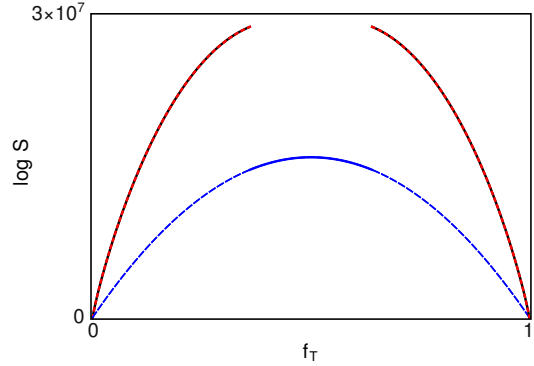}
\caption{\label{Fig5}State selection. Shown is the logarithm of the number of microscopic network configurations
realizing the different macroscopic states vs.\ the fraction of T nodes $\ft$, for a network with $N=10^4$ nodes,
estimated via moment closure approximation. The logarithm of the number of configurations is shown for branches
of non-bipartite steady state (black), unstable steady state (red, almost coinciding with the black) and the absorbing
bipartite state (blue). Solid lines denote states that are approached by the system.}
\end{figure}

Sustituting the steady states from the kinetic model~\ref{MCAsystem} yields the results shown in~\fref{Fig5}.
Outside the bipartite regime even the logarithm of the number of states is orders of magnitude larger in the
non-bipatrtite branches than in the bipartite branch. This suggests that the mechanism that drives the system
to the non-bipartite state, outside the bipartite regime, can be understood in terms of a \emph{configurational
entropy} which is maximized in the observed steady state. This state is also the one that minimizes the ratio
between the mean degree of the minority agents to that of the majority agents.

\section{Conclusions}
In the present paper, we proposed a toy model for nonlocal topological dynamics in
a simple network formation game. We showed that this model self-organizes to a completely
bipartite state in a wide parameter range, whereas links breaking the bipartiteness
appear in other parameter ranges. We explored the genesis of the bipartite regime
by network-level simulations, simple thermodynamics arguments, and a detailed kinetic
model.

The proposed system showed many characteristics that are closely reminiscent
of first-order phase transitions. It may therefore provide an analytically
tractable example of a discontinuous phase transition far from equilibrium.

Let us remark that, at present, we cannot conclusively prove that the proposed system
meets all criteria that are commonly applied to identify a phase transition. One concern
is perhaps that in the kinetic model, the phase transition does not show up as a single
discontinuous transition. Instead, the observed order parameter profile emerges due to
the presence of \emph{two} discontinuous transitions. However, we note that bifurcations,
unlike phase transitions, are not part of physical reality but features of a specific
model. The same physical transitions may therefore be described by different bifurcations
in different modeling frameworks.

We are confident that future works will confirm the nature of the transition proposed here.
A promising starting point for this work will be refinements of the kinetic theory. The kinetic
theory presented here is relatively simple and, in particular, neglects certain long-range
correlations. While the theory captures the behavior of the system relatively well, it is likely
that deeper insights can be gained by applying more sophisticated approximation schemes. In
particular, it is conceivable that this will change the bifurcation diagram turning the discontinuous
saddle-node bifurcations into continuous transcritical bifurcations and revealing the spinodal
branches of the system.

While more sophisticated approximation schemes, such as higher-order homogeneous approximations
or heterogeneous pair-approximations will require significantly more work, we believe that the
prospect of having an analytically tractable example of non-equilibrium first-order transitions
is well worth this effort.

\ack
The authors would like to gratefully acknowledge Güven Demirel for helpful
suggestions and careful review of the manuscript, and Veselina V.~Uzunova
for fruitful comments and stimulating discussions.

\end{document}